\documentclass[%
 aip,
rsi,%
 amsmath,amssymb,
 reprint,
longbibliography,
]{revtex4-1}

\usepackage[table]{xcolor}
\usepackage{sidecap}
\usepackage{graphicx}
\usepackage{dcolumn}
\usepackage{bm}
\usepackage{nicefrac}

\begin{document}

\preprint{AIP/123-QED}

\title[Rodr\'{i}guez et al.]{Boundary-induced effect on the spoke-like activity in E x B plasma}

\author{E. Rodr\'{i}guez}
 \altaffiliation[]{eduardor@princeton.edu}

\author{V. Skoutnev}

\author{Y. Raitses}%
 \altaffiliation{yraitses@pppl.gov}

\author{A. Powis}
\author{I. Kaganovich}%
\affiliation{%
Princeton Plasma Physics Laboratory, Princeton, New Jersey 08540, USA
}%
\author{A. Smolyakov}

\affiliation{
University of Saskachewan, Saskatoon, Saskatchewan S7N 5E2, Canada
}
\date{\today}

\begin{abstract}
The spoke instability in an $E\times B$ Penning discharge is shown to be strongly affected by the boundary that is perpendicular to $B$ field lines. The instability is the strongest when bounded by dielectric walls. With a conducting wall, biased to collect electron current from the plasma, the spoke becomes faster, less coherent and localised closer to the axis. The corresponding anomalous cross-field transport is assessed via simultaneous time-resolved measurements of plasma potential and density. This shows a dominant large-scale $E\times B$ anomalous character of the electron cross-field current for dielectric walls reaching 40--100\% of the discharge current, with an effective Hall parameter $\beta_\mathrm{eff}\sim10$. The anomalous current is greatly reduced with the conducting boundary (characterised by $\beta_\mathrm{eff}\sim10^2$). These experimental measurements are shown to be qualitatively consistent with the decrease of the $E$ field that triggers the collisionless Simon-Hoh instability.
\end{abstract}

\maketitle
\begin{quotation}
The following article has been submitted to Physics of Plasmas. After it is published, it will be found at https://aip.scitation.org/journal/php
\end{quotation}

\section{\label{sec:intro} Introduction:}

Cross-field discharges such as Hall thrusters\cite{morozovRev,boeuf17,Goebel08} for space applications\cite{lev18} and sputtering magnetrons\cite{swann88,kelly99} for material processing\cite{carcia03}, are susceptible to a number of performance-limiting unstable mechanisms\cite{zhurin99,lazurenko08,megaw34}. A better understanding of their physical origin thus proves essential.  \par
\begin{figure}
\includegraphics[width=0.45\textwidth]{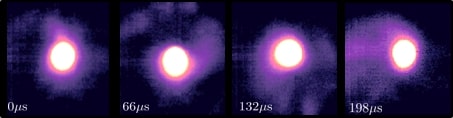}
\caption{Rotating spoke seen along the axis of the Penning device with a fast-frame camera. $B=80~$G is out of the page and $E$ is radially inwards (Xenon at 0.2~mTorr pressure).}\label{fig::spoke}
\end{figure}
Especially relevant, and the subject that is to concern us here, is the so called `spoke'\cite{McDonald11,Liu14,Ellison12,Anders14,Ehiasrian12} (see Figure \ref{fig::spoke}). This azimuthally rotating macroscopic structure of increased plasma density and enhanced ionisation is common to many $E\times B$ devices. Its rotation occurs quasi-coherently in the $E\times B$ direction, but with a speed slower than the $E\times B$ drift of magnetised electrons (typically by an order of magnitude). Moreover, the presence of the spoke has been associated in the literature with an increase in the electron cross-field mobility.\cite{Thomassen66,Janes66,Parker10,Ellison12,Pool15} This enhanced transport is significantly larger than those values expected from classical collisional calculations, and in this sense is deemed anomalous. This current has deleterious effects in, for example, Hall thrusters\cite{Janes62}, and is therefore a feature sought to be understood, controlled and ultimately eliminated. \par
\begin{figure}
\includegraphics[width=0.45\textwidth]{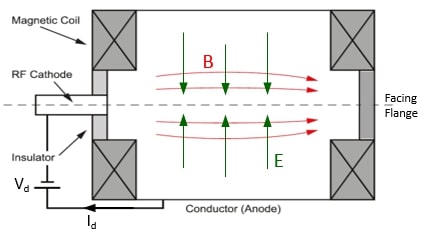}
\caption{Diagram showing a longitudinal cut of the Penning device including main features.}\label{fig::Penning}
\end{figure}

In the sake of understanding the origin of the spoke, many phenomenological accounts of potential sources, such as propagating ionisation fronts\cite{Janes66}, have been proposed. Studies of linear instabilities have also been carried out, leading to possible numerous growing modes\cite{Mikhail91,Dobro70,Litvak04,Smolya17,Sakawa93}. Amongst these, the Simon-Hoh instability\cite{Simon63,Hoh63}, originating from colinear $E$ field and density gradients, has drawn renewed attention. In particular its collisionless version \cite{Sakawa93} is driven exclusively by the electron $E\times B$ drift, and more complete considerations have also been recently proposed as part of general theory of gradient drift instabilities (GDI)\cite{Smolya17,Frias12,Frias13,MorinPoP2018}. \par 
Alongside these theoretical efforts, computational modelling has become increasingly important in extending present understanding of the spoke. Amongst the many numerical approaches available\cite{Boeuf13,Boeuf14,Carlsson18,Taccogna11}, simulations employing Particle-in-cell (PIC) codes\cite{Matyash13,Carlsson18} have lately gained relevance, including simulations of real sized devices\cite{Powis18}. Indeed, `spoke-like'  structures have been shown to appear in such recreated devices even in the simplest formulations, such as in the abscence of ionisation or collisions.\cite{Carlsson18,Powis18}  \par
	Imperative to the validity of all these existing predictions is, however, empirical evidence. This paper presents results of experimental studies of the $E\times B$ discharge shown in Figure \ref{fig::Penning} with a focus on the effects that changing boundary conditions have on the spoke activity. In particular, we show that the shortening of the $E\times B$ plasma by a metal wall as facing flange (see Fig. \ref{fig::Penning}) significantly reduces the spoke activity in part due to the reduction of the electric field, but does not suppress it completely. The shortening of the $E\times B$ plasma with the conductive wall is similar to the well-known Simon's effect\cite{Simon55}, but its effect on the spoke is the finding of this paper. Indeed, a different type of azimuthal activity with twice and three times higher frequencies than typical spoke oscillations, and modified spatial distribution is observed for the same operating conditions as in the $E\times B$ plasma bounded with dielectric walls as facing flange. An enhanced cross-field electron conduction due to perturbations by the spoke is also studied and significantly reduced. A final comparison to some predictions of the collisionless Simon-Hoh instability (CSHI) is also presented. \par

\section{Experiment background}

\subsection{\label{sec:theo}Theoretical basis}

For an analysis of experimental results and their comparison with theoretical predictions, collisionless Simon-Hoh instability (CSHI) is considered here, as developed originally in [\onlinecite{Sakawa93}] and obtained as a limiting form of GDI in Refs [\onlinecite{Frias12},\onlinecite{Frias13}]. \par
The classical CSHI theory invokes the following assumptions. First of all, the approach is linear and is developed for an infinte slab geometry, with one periodic coordinate and a uniform direction. Secondly, both electrons and ions are treated as collisionless fluids but, the plasma is only partially magnetised. Then, magnetised electrons have significant $E\times B$ and diamagnetic drifts. Finally, and particular to the Penning discharge, the magnetic field gradients are neglected in comparison to potential or density ones. \par 
This fluid mode is dominated by the radial electric field, $E_r$, and density gradient length scale, $L^{-1}_n\equiv\nicefrac{|\nabla n|}{n}$. The form of the dispersion is given by\cite{Sakawa93}
\begin{equation}
\frac{\omega_*}{\omega-\omega_0}=\frac{k_\theta^2c_s^2}{\omega^2}
\label{eqn::disMSH}
\end{equation}
where, $\omega_*=\nicefrac{- k_\theta k_BT_e}{eBL_n}$ is the electron diamagnetic drift frequency, $\omega_0=k_\theta\nicefrac{E_r}{B}$ is the $E\times B$ drift frequency, $c_s$ is the ion sound speed, and $k_\theta$ is the azimuthal wavenumber. \par
From Equation (\ref{eqn::disMSH}), the growth rate $\gamma$ of the instability is, \par
\begin{equation}
\gamma = \frac{k_\theta c_s}{\omega_*} \sqrt{\omega_0\omega_*-\frac{k_\theta^2c_s^2}{4}}. \label{eqn::grwth}
\end{equation} 
It is readily seen that for the system to be unstable (ie. $\gamma$ real) collinearity of $E_r$ and the density gradients is necessary, as it is the case in these experiments (see Appendix B). \par

\subsection{\label{sec:exper}Experimental set-up}
The entirety of the experiments presented in this paper were done in the $E\times B$ Penning setup shown in Figure \ref{fig::Penning} and described elsewhere.\cite{Raitses15} The simplicity of this experiment makes it an attractive case of study, still bearing common ground with other larger efforts with similar geometry such as MISTRAL\cite{Anna11} and LAPD of UCLA\cite{Gekel91}.\par
  In this setup, a partially magnetized plasma with electron temperature and density of $T_e\sim1-5~$eV and $n\sim 10^{16}-10^{17}$m$^{-3}$, respectively, is generated in a 26 cm diameter ($D$) and 50 cm length ($L$) non-magnetic, stainless steel, 6-way cross chamber equipped with a turbo-molecular pump and a mechanical pump. With Xenon as the working gas, a typical operating pressure in the chamber during the described experiments is 0.2~mTorr. The plasma is produced by impact ionization of Xenon atoms with energetic electrons extracted from the RF-plasma cathode.\cite{Rait09} The cathode is placed at one of the ports of the vacuum chamber to enable the injection of electrons along the applied magnetic field. A typical electron extracting voltage of $V_d\sim55$~V was applied between a metal wall of the RF cathode and the chamber (ground). Thus, the chamber acts as the anode with respect to the RF cathode. Electrons from the cathode flow to the chamber along a solenoidal-type magnetic field of $B=30-150$~G produced by a set of coils placed outside the chamber. A radial $E$ field is also present, with magnitudes $E\lesssim 200$Vm$^{-1}$. In the described experiments, the extraction current (equal to the DC discharge current) was liimited to $I_d\sim1.2$~A.  Under typical operating conditions, electrons are magnetized with electron gyroradius of $\rho_e\sim0.3-1.7~$mm$\ll D, L$, while ions are on average weakly to non-magnetized ($\rho_i\sim 0.3-1.5~$m). \par

\par
In order to study the effect of boundary conditions on the spoke and the electron cross-field current, the Penning setup was operated in two configurations. In one configuration (Case I), the chamber walls perpendicular to the magnetic field lines (see facing flange in Fig. \ref{fig::Penning}) were dielectric, made from either machinable glass ceramic or Pyrex. The transparent Pyrex was used to allow monitoring the spoke activity with a fast framing camera (see Figure \ref{fig::spoke}). There were no observable differences in the plasma operation of the Penning discharge between these two dielectric materials. \par

In the other configuration of the Penning discharge (Case II), a non-magnetic stainless steel flange was used as a magnetic field facing wall. The wall was electrically connected to the chamber and, thereby, it also acted as anode with respect to the RF cathode. In this set-up, the electron flow from the cathode is anticipated to be directly collected along the magnetic field lines connecting the cathode and the flange (Fig. \ref{fig::Penning}).\cite{Simon55} It is also expected that, under such conditions, the plasma confined within the flux tube connecting cathode and anode will be close to equipotentiality. Then, in the absence of the electric field across the magnetic field, Eq. \ref{eqn::grwth} predicts no CSHI instability. 
In this context, a reduced spoke activity would be consistent with CSHI, and could be taken as indication that CSHI likely plays a role in the spoke formation. The purpose of experiments in this Penning discharge configuration is, in part, to test this prediction.

\subsection{\label{sec:measuring}Plasma diagnostics}

\subsubsection{\label{sec:density}Plasma density ($n$) -- ion probes}
A negatively biased ion probe is used to deduce plasma density, $n$. 
The probe collects saturated current, $I_{\mathrm{ion}}$, in the Bohm regime\cite{Merlino07}, without significant sheath expansion. The lack of expansion was experimentally resolved by observing collection current variations when changing probe negative bias; this gave changes below 5\% per 10~V. A 10~k$\Omega$ shunt resistor was used to measure the current in the probe circuit. Thus, from the voltage drop measured across the shunt, the ion density is given as
\begin{equation}
n=\frac{I_{\mathrm{ion}}}{\gamma eA_pc_s} \label{eqn::ionPrb}
\end{equation}
where $c_s=\sqrt{\nicefrac{k_BT_e}{M}}$ is the Bohm velocity and $A_p$ is the probe area. \par 
The factor $\gamma$ corresponds to a geometry dependent, effective presheath density correction from radial motion limited theory\cite{Chen02}. A constant approximated value of $\gamma\sim1$ is taken for the investigated plasma conditions. \par
In the right hand side of Equation (\ref{eqn::ionPrb}), a constant value is also taken for the electron temperature of the plasma, which will, as a result, be a source of uncertainty. Spatial and uncorrelated temporal\cite{Val18} variations not accounted for will lead, respectively, to systematic and random errors on the final value of $n$ (see Appendix A). Note that non-Maxwellian features of the EEDF could also be potentially important, but comparison to sweeping bias Langmuir probe measurements indicated a limited disagreement to below $\sim10$\%.

\subsubsection{\label{sec:Vp}Plasma potential ($V_p$) -- floating emissive probe}
A DC heated, floating emissive probe made from a 0.1~mm diameter thoriated tungsten wire is used to measure the plasma potential, $V_p$. The probe is operated in the regime of strong thermionic electron emission, manifested itself as the saturation of the hot-probe floating potential with respect to the ground.\cite{Sheehan17} For the correct interpretation of this floating potential reading, the contribution from the heating voltage ($V_h$), typically $V_h\sim 4~$V, ought to be taken into consideration. Indeed, $\nicefrac{1}{2}V_h$ is subtracted from the measured voltage value.\cite{Mrav90} In general, $V_p$ is related to the corrected measured probe voltage, $V_f^\mathrm{hot}$, by
\begin{equation}
	 V_p = V_f^\mathrm{hot}+\alpha T_e \label{eqn::Brian}
\end{equation}
where $\alpha$ is some constant $O(1)$ that depends on the particularities of the plasma under study\cite{Kraus18}. \par
To assess how good an approximation $V_p\approx V_f^\mathrm{hot}$ is, the smallness of the term $\alpha T_e$ in Eq. (\ref{eqn::Brian}) was tested by comparing measurements to a sweeping Langmuir probe in the same discharge and at the same spatial location, using the second derivative of the measured IV traces to find the plasma potential\cite{God11,Druv40} (see Appendix A).  

\par

\subsubsection{\label{sec:j}Cross-field current ($j_{\perp,E\mathrm{x}B}$) -- two-probe method and rotating-wave approximation}
The anomalous electron cross-field current resulting from fluctuations of local electric fields and plasma density is measured indirectly. For this measurements, the transport is assumed to be dominated by $E_\theta\times B$ variations, so that \par
\begin{equation}
	j_{\perp,E\mathrm{x}B}(r,\theta)=n(r,\theta)e\frac{E_\theta(r,\theta)}{B} \label{eqn::anomCurr}
\end{equation}
The form of Eq. (\ref{eqn::anomCurr}) suggests that, for a net radial electron transport (i.e. a non-vanishing azimuthal averge) to exist, $E_\theta$ and $n$ should have a small relative phase.  In light of this, local measurements of density and azimuthal electric field are required. The two-probe method does so by using, simultaneously, an ion probe and an emissive probe, described in the two preceeding sub-sections. These two distinct probes record $n$ and $V_p$ respectively from poloidaly closely separated locations (a $\sim 0.3~$cm apart) without significant shadowing. \par

In order to obtain $E_\theta$ from these measurements, the recorded $V_p$ time series needs to be mapped onto the spatial domain, where the definition of $E_{\theta}\equiv-\nabla_\theta V_p$ may be applied. An approximated projection is possible under the assumption of what we call the \textbf{rotating wave approximation}. This approximation scheme exploits the quasi-periodic rotation of the spoke, which is taken to be rigidly and uniformly whirling about the axis. In that case, the linear map is trivial, but gives rise to 
	an uncertainty of up to $\sim70$\% (see Appendix A). It is noted here that small turbulent scales are not resolved in this approach. The uncertainty does however not preclude the main conclusions in this paper. 
 This will however be the dominant source of uncertainty in $j_{\perp,E\times B}$.

\section{Spoke characterisation}
\subsection{Spoke activity reduction with metallic flange} 
\begin{figure}
\hspace*{-0.3cm}
\includegraphics[width=0.5\textwidth]{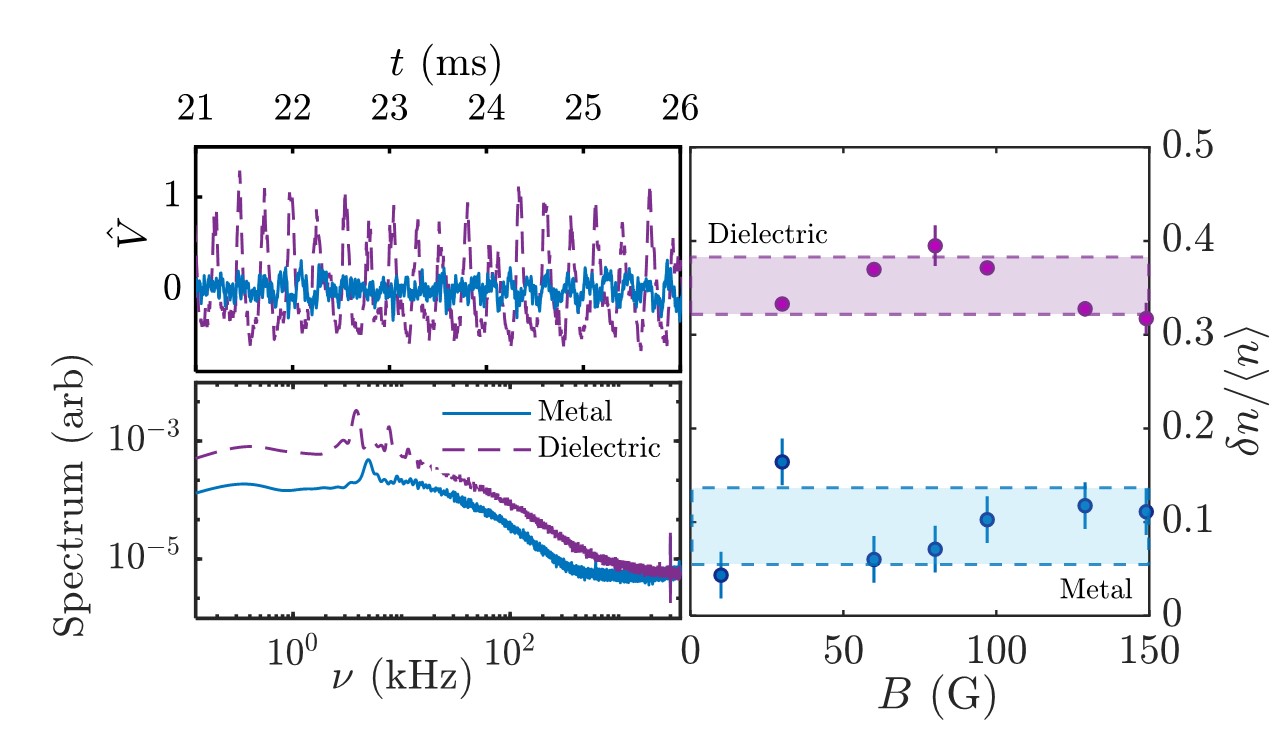}
\caption{Change in oscillatory behaviour of the plasma when changing the boundary. \textbf{(LEFT)} Plots of normalised ion probe signal $\hat{V}=\frac{n-\langle n\rangle}{\langle n\rangle}$ for metal and glass boundaries, and their corresponding spectrum, for $B=30~$G. \textbf{(RIGHT)} Magnitude of variations, representative of instability, defined as $\delta n/\langle n \rangle$ as a function of $B$; $\delta n$ is the standard deviation of the time signal of the main spectral component of the perturbation.}\label{fig::fig2}
\end{figure}

The configuration in Case II led to a clear reduction in spoke activity. This improvement in stability is described by a figure of merit defined as $\hat{V}=\frac{n-\langle n\rangle}{\langle n\rangle}$. That is to say, the relative size of the recorded density fluctuations is used as proxy for instability. An example of the reduced activity and the magnitudes for numerous $B$ fields are shown in Figure \ref{fig::fig2}. \par
The density variations (see right panel of Figure \ref{fig::fig2}) are reduced by a factor of $\sim3.4(3)$ on average, over a range $B=$10--150~G. This suggests that the stabilisation mechanism introduced is not just incidental, but holds for an extended set of parameters. \par
Nevertheless, the quieter plasma obtained with the metal does still show a prominent peak in its spectrum. The properties of this prevailing mode will now be compared to the spoke in Case I.

\subsection{Spoke changes due to boundary}
\subsubsection{Changes in shape and frequency}
\begin{figure}
\hspace*{-0.6cm}
\includegraphics[width=0.4\textwidth]{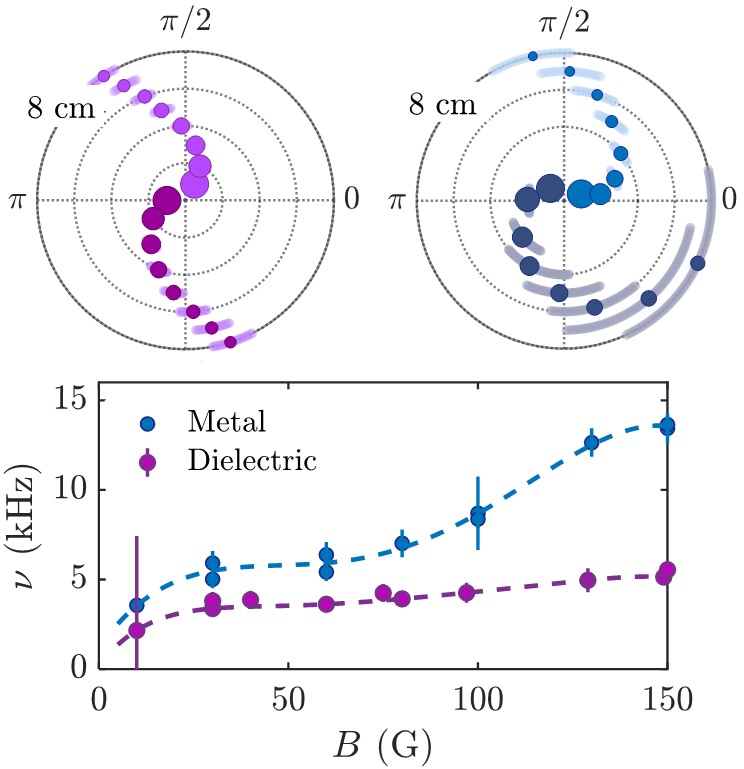}
\caption{(LOWER) Rotation frequency $\nu=\nu(B)$ for metal and dielectric boundaries, showing the larger frequency in the metallic case. (UPPER) Spoke shape for $B=30~$G, each of the two arms in each plot corresponding to opposite directions of $B$. The shade represents the central  68\% about the median. Comparing the shades at each radii of the lower arm in the right and left shapes respectively, a roughly 4-fold difference is observed. Besides, the dispersion at $r=8~$cm of the right arm is of the order of $\nicefrac{\pi}{2}$. }\label{fig::fig3}
\end{figure}

The most basic characteristics of a typical spoke in a Penning discharge are its slower frequency as compared to the $E\times B$ drift, its $m=1$ mode nature and shape. The changes in these properties are exemplified in Figure \ref{fig::fig3}, and suggest that the spoke is of a different nature. \par

The shapes of the spokes in the top panels have been reconstructed based on data from simultaneous measurements of two ion probes, separated poloidally by an angle of $\nicefrac{\pi}{2}$. The radial position of one of the probes was fixed at $r\sim6$~cm, while the other changed so as to sample the radial extent of the spoke. The cross-correlation phase between the signals is the quantity shown in each of the polar plots, each arm within a diagram representing an opposite $B$ direction. \par
These graphs show that the rotation in the $E\times B$ direction persists when the boundary is modified. However, a n approximately 4-fold increase in the dispersion of the correlation phase in Case II suggests that the spoke has lost coherency and/or rigidity. In addition, for Case II, a significant loss of correlation at larger radii may be observed, reaching phase deviations of up to $\Delta\phi\sim\nicefrac{\pi}{2}~$rad for the outermost locations. This observation may be linked to the spoke being located closer to the axis. \par
Regarding changes in the frequency of rotation (see lower panel), the modified oscillations occur approximately twice as fast, closer to the characteristic $E\times B$ drift frequency, which falls by a factor order $\sim2$. As a way of example, for $B=150~$G with the metal wall,
\begin{equation*}
\nu_{E\times B}=\frac{E_r}{2\pi rB}\approx\frac{50\mathrm{Vm}^{-1}}{2\pi\times4\mathrm{cm}\times150\mathrm{G}}\approx 13~\mathrm{kHz,}
\end{equation*}
while for glass $\nu_{E\times B}\approx43~$kHz. The corresponding spoke frequencies are 14~kHz and 6~kHz respectively. The increased magnitude of $\nu$ is also accompanied by an increased complexity in $\nu(B)$, with the possibility of various competing modes, something to be considered in future work. \par

\subsubsection{Changes in anomalous $E\times B$ electron cross-field current}
\begin{figure}
\hspace*{-01.25cm}
\includegraphics[width=0.55\textwidth]{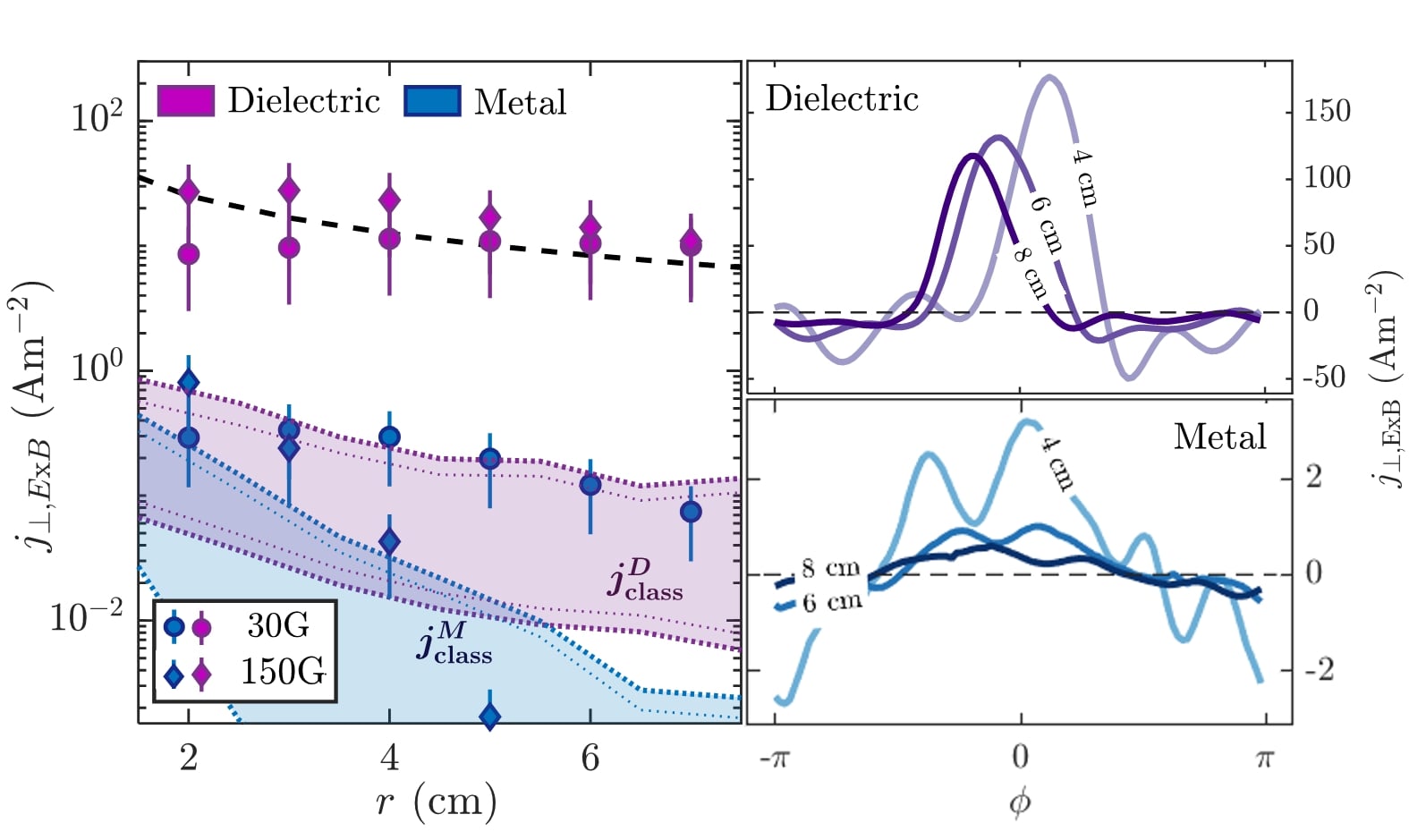}
\caption{(LEFT) Comparison of time averaged cross field anomalous current density with radius for metal and glass boundaries, shown for $B=30,150~$G. The shaded regions represent classical transport estimates given measured gradients for values of $B$ spanning 30--150~G. The broken black curve corresponds to a uniform distribution of $I_d$, i.e. $j=I_d/2\pi rL$ (RIGHT) Spoke phase resolved transport, with $\phi=0$ corresponding to the density maximum ($B=30$~G)}\label{fig::fig4}
\end{figure}

The anomalous electron cross-field current, $j_{\perp,E\mathrm{x}B}$, in the radial direction is investigated using the described two-probe method. Figure \ref{fig::fig4} shows the average magnitude of $j_{\perp,E\mathrm{x}B}$ and clasical transport values, as well as the relation of $j_{\perp,E\mathrm{x}B}$ to the passage of the spoke. The figure also compares the measured cross field current with a cross field current estimate under the assumption that this is uniformly distributed in axial and azimuthal directions of the Penning discharge, i.e. $I_d/2\pi rL$. \par

The current distribution as a function of the spoke phase shows for Case I (i.e. dielectric) a close relation between the $E\times B$ induced transport and the spoke. This correspondance appears as a narrow (i.e. FWHM of $\Delta\phi\sim\nicefrac{\pi}{4}-\nicefrac{\pi}{3}$) conduction channel in Figure \ref{fig::fig4}. From the average profiles, it follows that this $E\times B$ current is the main contributor to the total discharge current; indeed, measurements lie within agreement (i.e. relative value of 40--100\%) with the uniform distribution estimates. These observations of $E\times B$ transport are consistent with results from previous experiments performed in the same Penning discharge using a fast sweeping Langmuir probe\cite{Val18} and Hall thrusters\cite{McDon11,Ellison12}.  
\par

In addition, $j_{\perp,E\mathrm{x}B}$ is soundly anomalous, given that it lies around two orders of magnitude over the classical electron-neutral collisional transport levels (eg. for $B=30~$G and $r=8~$cm, $j_\mathrm{anom}/j_\mathrm{class}\approx90(70)$). The classical transport may be estimated using experimentally measured field gradients in
\begin{equation}
	j^{\mathrm{class}}_r=\frac{\sigma}{1+(\omega_{ce}/\nu_{en})}\left(E_r+\frac{\partial T_e}{\partial r}+T_e\frac{\partial \ln n}{\partial r}\right)
\end{equation}
where
$$ \sigma=\frac{ne^2}{m_e\nu_{en}} $$
and
$$  \nu_{en}=n_0\langle \sigma_{en} v\rangle $$ , and $n_0\approx 1.2\times10^{19}~$m$^{-3}$ is the neutral density, $\omega_{ce}$ is the electron cyclotron frequency, $\sigma_{en}$ is the electron-neutral collisional cross-section and $v$ is the speed of the electron species. \par
 Importantly, it is here emphasised that the large scale spoke disturbances are seen to be the primary responsibles for the majority of the transport (40--100\%). Given the magnitude of the classical contribution, the remaining of the transport could be associated to other processes mediated by smaller scale\cite{Powis18}. This matter requires additional research to include measurements of smaller scale.  \par
The large observed anomality may be also characterised by the effective Hall parameter defined as $\beta^i_\mathrm{eff}=\omega_{ce}/\nu_\mathrm{eff}$, where $\nu_\mathrm{eff}$ is the effective collisional frequency that gives the observed current and $i$ is a superscript that labels the two dielectric (D) and metal (M) cases. This non-dimensional parameter is approximatedly $\beta^D_\mathrm{eff}\approx8(6)$ for $B=30~$G in Case I at the edge $r=8~$cm, consistent with values presented in some numerical PIC simulations of Penning discharges.\cite{Powis18,Carlsson18} \par
When the metallic flange is placed, the total cross field transport in the plasma drops very significantly (see Fig. \ref{fig::fig4}), as expected from the short circuit effect. The anomalous character of transport, although still relevant, does also fall, as $j_{\perp,E\mathrm{x}B}$ values are closer to classical collisional transport values (eg. $\sim6$ times on average for $B=30~$G cases, with $\beta^M_\mathrm{eff}\approx220(150)$). At the same time, measured current is less correlated with the passage of the spoke; indeed, current profiles become spatially broader ($\Delta\phi\sim\nicefrac{2\pi}{3}$), noisier and less prominent. This observation evidences the relation between transport and the spoke. \par

In conclusion, the reduction of the spoke activity appears connected to the drop in the total and anomalous cross field transports.

\section{Comparison with Theory}
\begin{figure}
\hspace*{-0.2cm}
\includegraphics[width=0.5\textwidth]{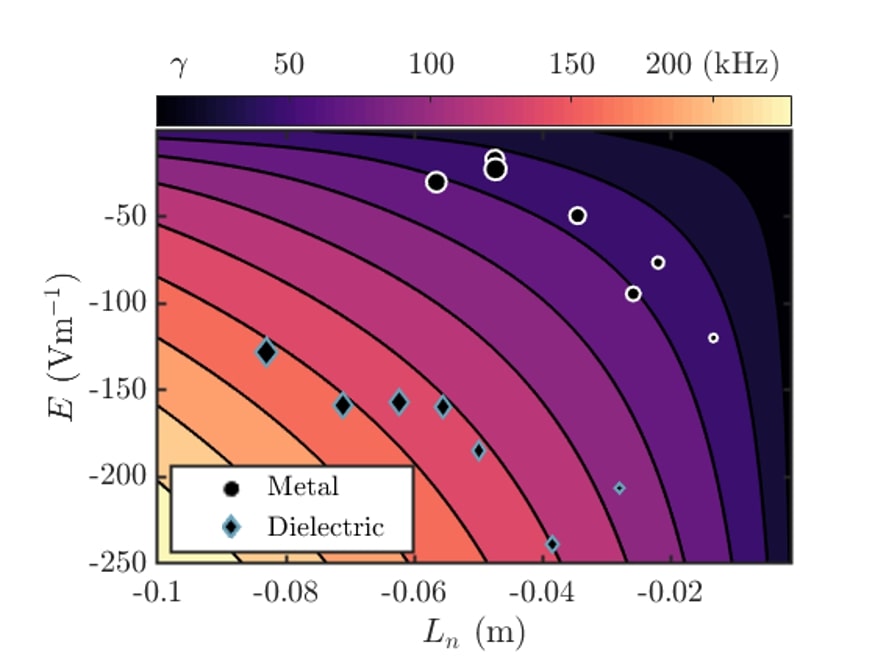}
\caption{Collisionless Simon-Hoh instability growth rate, $\gamma$, in $E-L_n$ parameter space. The scattered data represents mean measured gradients for $B=150~$G, with the increasing symbol size representing larger radii measurements for 1--8~cm. 
 }\label{fig::fig5}
\end{figure}

The observed improvement in stability is now compared to the growth rate predictions of CSHI. To determine $\gamma$ associated to the experimental conditions, mean time averaged radial profiles of $V_p$ and $n$ are measured empirically (see Appendix B). 
 Taking $k_\theta\sim\nicefrac{1}{r}$, where $r$ represents radial position, all information needed for Eq. (\ref{eqn::grwth}) is available. Figure \ref{fig::fig5} shows as scatter data the measured quantities in ($E$,$L_n$) space, for $B=150~$G. Every pair of coordinates in this space is indeed associated to a growth rate, and this is represented as a coloured contour. \par

According to CSHI theory, gradients are consistent with the improved stability; the gradients corresponding to the dielectric caselying in higher $\gamma$ value. This prediction, though in qualitative agreement with experiment, presents a number of limitations that further theoretical work should amend. \par
To start with, it applies field gradient measurements from the non-linear saturated regime to the linear theory formalism. 
It could be argued that, assuming that on average, due to the forcing and dissipation balance between the electron injection and the instability, the spoke should not change radial gradients dramatically, the qualitative linear description could be expected to be approximatedly correct. Perhaps more clearly a failure of the linear treatment of CSHI is that it does not predict the observed prominency of the $m=1$ mode, instead it suggests that $m>1$ are more unstable. In recent PIC simulations of the Penning discharge, the dominance of m=1 mode was also observed\cite{Powis18,Carlsson18}. \par
Additional limitations, especially describing the metal case, come from disregarding several physically relevant features. For instance, due to the decrease of radial $E$, additional gradients (such as $B$ or $T_e$) could gain relevance and need to be considered, e.g. using the more complete form of GDI\cite{Smolya17,Frias12,Frias13}. Even with such an extension, relevant features of the system take us further away from classical CSHI: axial uniformity may break down (there is a significant net axial current flow), the short circuit effect\cite{Simon55} is not acccounted for and possibly significant sheaths\cite{MorinPoP2018,SmolyakovPRL2013} may be present. A more detailed numerical, theoretical and experimental exploration is left for future work. \par
Nonetheless, the qualitatively consistent behaviour with the electric field and density gradient trends following from CSHI suggest as a possible theoretical framework the more general GDI theory, perhaps including additional effects such as the sheath.  

\section{Conclusion}

The spoke instability sustained in a low-density plasma in an $E\times B$ Penning discharge has been demonstrated to be significantly reduced by changing boundary properties. The introduction of an all-metallic surface connected to the anode in front of the flow of electrons from the cathode, led to a reduction in the prominence of fluctuations, as hypothesised. However, a modified saturated structure persisted, characterised by a faster, more radially confined and less coherent rotation than that of the spoke with the dielectric. 
 \par
The cross field anomalous $E\times B$ electron transport due to the large-scale spoke perturbations was measured in-situ and, for the dielectric, seen to account for a $40-100$\% of the total discharge current, resulting in a $\beta^D_\mathrm{eff}\approx10$ consistent with computational work\cite{Powis18,Carlsson18}. Transport becomes more spread and closer to levels of classical collisional transport with the metal as boundary ($\beta^M_\mathrm{eff}\approx10\beta^D_\mathrm{eff}$). 
These changes in transport imply an inhibited radial loss of electrons, which could be of relevance in design of devices such as Hall Thrusters (more specifically, anode layer Hall thruster, Camila Hall thruster\cite{Kron12,Kapul11}, segmented Hall thruster\cite{Fisch01,Diam06}, magnetically shielded Hall thruster\cite{Conver17}) and sputtering magnetrons\cite{Anders14,Ito15}. \par
Finally, the observations were shown to be generally consistent with the electric field and density gradient trends following from the standard collisionless Simon-Hoh instability. 
However, additional gradients\cite{Smolya17,Frias12,Frias13}, sheath effects\cite{MorinPoP2018,SmolyakovPRL2013} or significant axial non-uniformities are recognised as likely relevant aspects especially in the metallic case, and should be considered in a more comprehensive framework. The investigation of these added complexities is left for future numerical, experimental and theoretical work. 

\begin{acknowledgments}
Thanks to Brian Kraus, Andy Alt and Johan Carlson for fruitful discussions. 

This work was supported by the Air Force Office of Scientific Research (AFOSR).
\end{acknowledgments}
\appendix

\section{Experimental uncertainty}
In this appendix the uncertainty sources taken into consideration for the measurements in this paper are spelled out for completeness.
\begin{enumerate}
\item \textbf{Plasma density (n)}:\par
See Table \ref{tab::dens} for the uncertainty contribution to the plasma density $n$ measurement with ion probes. The first two entries (measured radial variation in temperature $T_e(r)$ based on floating cold and emissive probe\cite{Sheehan17} and calculated correction parameter\cite{Chen02} $\gamma(r)$ ) are anti-correlated and result from misregarding spatial variations. The remaining represent uncorrelated sources such as: time fluctuations\cite{Val18} of $T_e(t)$, theoretical indeterminacy\cite{Yev05} in temperature measurement ($\Delta T_\mathrm{TH}$) and correction for sheath expansion from the measured saturation current changes with bias voltage ($\Delta n_\mathrm{SAT}$). The total uncertainty is obtained by combining all these various sources of indeterminacy.
\begin{SCtable}[2][h]
\hspace*{-0.4cm}
\vspace*{0.5cm}
\begin{tabular}{|c|c|}\hline
    Source & $\Delta n/n$ \\\hline \hline
    $\gamma(r)$ & 15\% \\ 
    $T_e(r)$ & 25\% \\ \hline
    $T_e(t)$ & 5\% \\ 
    $\Delta T_\mathrm{TH}$ & 10\% \\ 
    $\Delta n_\mathrm{SAT}$ & 10\% \\ \hline
    $\Delta n_\mathrm{TOT}$ & 20\% \\ \hline
\end{tabular}
\hspace*{0.5cm}
\caption{Uncertainty sources to the plasma density $n$: radial variation in temperature $T_e(r)$, calculated correction parameter\cite{Chen02} $\gamma(r)$, time fluctuations\cite{Val18} of $T_e(t)$, theoretical indeterminacy\cite{Yev05} in temperature measurement ($\Delta T_\mathrm{TH}$) and correction for sheath expansion ($\Delta n_\mathrm{SAT}$). The total uncertainty results from the statistical combination of all. } \label{tab::dens}
\end{SCtable}

\item \textbf{Plasma potential ($V_p$)}: \par
See Table \ref{tab::pot} for the uncertainty contribution to the plasma potential $V_p$ measurement with the emissive probe. The first entry correponds to the error due to ignoring the term $\alpha T_e$ in Eq. \ref{eqn::Brian}, found to be $-\alpha\sim0.1-0.2$ and $T_e\sim2~$eV. This itself  has a randomly fluctuating part\cite{Val18} to it $\sim5$\%. The uncertainty in the heating voltage $V_h$ measurement also contributes, even assuming the theoretical correctness of the $\nicefrac{1}{2}V_h$ correction.\cite{Mrav90}
\begin{SCtable}[2][h]
\hspace*{-0.4cm}
\vspace*{0.5cm}
\begin{tabular}{|c|c|}\hline
    Source & $\Delta V_p~$(V) \\\hline \hline
    $\alpha T_e$ & 0.2-0.4 \\ 
    $V_\mathrm{h}$ & 0.1 \\ \hline
    $\Delta V_{p,\mathrm{TOT}}$ & 0.2-0.4 \\ \hline
\end{tabular}
\hspace*{0.5cm}
\caption{Uncertainty sources to the plasma potential $V_p$: from ignoring the term $\alpha T_e$ in Eq. \ref{eqn::Brian}, and the heating voltage $V_h$.} \label{tab::pot}
\end{SCtable}
%\vspace*{2.5cm}

\item \textbf{Electron cross-field current ($j_{\perp,E\mathrm{x}B}$)}: \par
See Table \ref{tab::curr} fot the uncertainty contribution to the electron cross-field current $j_{\perp,E\mathrm{x}B}$ measurement with the two-probe method. The first two entries correpond to the error in determining $E_\theta$: due to fluctuations in $\alpha T_e$ that modify the actual $V_p$, and local variations in the rotation speed $\omega_\mathrm{rot}$, whose uniformity was assumed in the rotating wave approximation. The $\omega_\mathrm{rot}$ variations are assessed experimentally using correlation of signals from two closely separated identical ion probes. Related to $n$, an overall scaling uncertainty $\Delta n$ is to be considered. Also, some error arises due to non-coincident measurements of $n$ and $E_\theta$; this is estimated considering a displacement assuming a sinusoidal $n(\theta)$ profile.
\begin{SCtable}[2][h]
\hspace*{-0.5cm}
\vspace*{0.5cm}
\begin{tabular}{|c|c|c|}\hline
     & Source & $\Delta j/j$ \\\hline \hline
    $E_\theta$ & $\Delta(\alpha T_e)$ & 1\% \\ 
    & $\omega_\mathrm{rot}$ & 70\% \\ \hline
    $n$ & $\Delta n$ & 20\% \\
	& $n(\Delta\theta)$ & 5\% \\ \hline
    $\Delta j_\mathrm{tot}$ & --  & 70\% \\ \hline
\end{tabular}
\hspace*{0.5cm}
\caption{Uncertainty contribution to the electron cross-field current $j_{\perp,E\mathrm{x}B}$: fluctuations in $\alpha T_e$, local variations in rotation speed $\omega_\mathrm{rot}$, overall scaling uncertainty $\Delta n$, non-coincident measurements of $n$ and $E_\theta$.} \label{tab::curr}
\end{SCtable}

\end{enumerate}

\section{Field gradient sample}
\begin{figure}[h]
\includegraphics[width=0.5\textwidth]{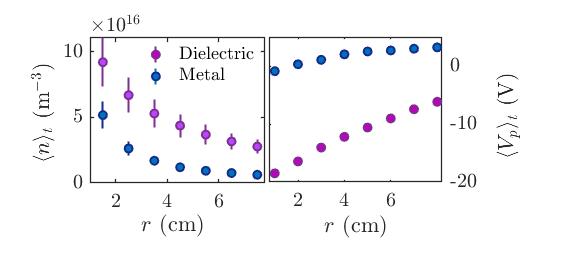}
\vspace*{-1cm}
\caption{Density and plasma potential profiles for $B=150G$ for both metallic and dielectric boundary cases.}\label{fig::fields}
\end{figure}

An example of what the field distribution looks like is here presented in Fig. \ref{fig::fields} for both Cases I and II for $B=150~$G.

\nocite{*}
\bibliography{main}% Produces the bibliography via BibTeX.

\end{document}